\documentclass[a4paper,twoside]{article}

\usepackage{epsfig}
\usepackage{subfigure}
\usepackage{calc}
\usepackage{amssymb}
\usepackage{amstext}
\usepackage{amsmath}
\usepackage{amsthm}
\usepackage{dsfont}
\usepackage{multicol}
\usepackage{pslatex}
\usepackage{SCITEPRESS}
\usepackage[small]{caption}
\usepackage[square]{natbib}
\usepackage{url}
\usepackage[pdfpagelabels,colorlinks,urlcolor=blue,linkcolor=blue,citecolor=blue]{hyperref}
\hypersetup{pageanchor=false,
    bookmarksnumbered=true
}

\newcommand{\RR}{{{\rm I \kern -0.2em R}}}
\newcommand{\DD}{{{\rm I \kern -0.2em D}}}
\newcommand{\CC}{{{\mbox{\rm \hspace*{0.05ex}
\rule[.18ex]{.18ex}{1.24ex} \kern -.65em C}}}}
\newcommand{\be}{\begin{equation}}
\newcommand{\ee}{\end{equation}}
\newcommand{\ba}{\left [ \begin{array}}
\newcommand{\ea}{\end{array} \right ]}

\newcommand{\rank}{\mathop{\mathrm{rank}}}

\newtheorem{theorem}{Theorem}
\newtheorem{corollary}{Corollary}

\begin{document}

\title{	
Fault Detection and Diagnosis: Computational Issues and Tools}

\author{\authorname{Andreas Varga }
\affiliation{Gilching, Germany
}
\email{varga.andreas@gmail.com}
}

\onecolumn \maketitle \normalsize \vfill
\section*{Abstract}
{A representative set of fault diagnosis problems is formulated for linear time-invariant systems with additive faults.  For all formulated problems, general existence conditions of their solutions are given. An  overview of  recent developments of computational methods for the synthesis of fault detection filters is presented and available software tools are described. }

\section*{Keywords}
{Fault detection and diagnosis; Fault detection and isolation; Model-matching synthesis; Numerical methods; Software tools} \\

\subsubsection*{AMS subject classifications:} 93-04, 93B40, 93B50, 93B55, 93C05, 93C15, 93C35, 93C55

\section{Introduction}
The theoretical developments of the model-based fault
detection and diagnosis  for linear time-invariant systems are essentially completed and widely  documented in several monographs and textbooks \citep{Gert98,Chen99,Iser06,Ding13,Blan16,Varg17}. Also the development of numerically reliable computational procedures for the synthesis of fault detection filters has been mostly finalized in the last decade. The author's book \citep{Varg17} is mainly dedicated to the presentation of the evolved new generation of computational synthesis procedures. These procedures are well suited as the basis of implementation of versatile software tools, which allow the solution of synthesis problems in the most general setting using the best available numerical methods.

This article presents an overview of the recent developments both in the computational synthesis procedures and associated software tools.
We start by formulating a canonical set of exact and approximate synthesis problems of fault detection filters and give general solvability conditions, which guarantee the existence of the solutions in the most general setting, for both continuous- and discrete-time systems. For each of the formulated problems, general synthesis procedures have been developed in \citep{Varg17}, which guarantee the determination of a solution to a specific problem whenever a solution exists. The development of these procedures relies on several computational paradigms, which are discussed in details. They underly efficient and numerically reliable computational methods, which served for the implementation of  software tools for the analysis of fault diagnosis problems and synthesis of fault detection filters.

\section{Plant Models with Additive Faults}
The underlying plant models to the discussed synthesis methods (also called \emph{synthesis models}) are linear time-invariant (LTI) system models, where the faults are equated with special (unknown) disturbance inputs.
An important class of models with additive faults arises when defining the
fault signals for two main categories of faults, namely, actuator and sensor faults. Two basic forms of synthesis models are used.

The input-output plant model with additive faults  has the form
\begin{align} \label{systemw}
\hspace*{-5mm}{\mathbf{y}}(\lambda) = & \;\;
G_u(\lambda){\mathbf{u}}(\lambda) +
G_d(\lambda){\mathbf{d}}(\lambda) + \nonumber \\ & \;\;
G_f(\lambda){\mathbf{f}}(\lambda) +
G_w(\lambda){\mathbf{w}}(\lambda)
,
\end{align}
where  ${\mathbf{y}}(\lambda)$, ${\mathbf{u}}(\lambda)$,
${\mathbf{d}}(\lambda)$, ${\mathbf{f}}(\lambda)$, and
${\mathbf{w}}(\lambda)$, with boldface notation, are the Laplace-transformed (in the continuous-time case) or Z-transformed (in the discrete-time case)  $p$-dimensional system output vector $y(t)$,
$m_u$-dimensional control input vector $u(t)$,
$m_d$-dimensional disturbance vector $d(t)$,
$m_f$-dimensional fault vector $f(t)$ and
$m_w$-dimensional noise vector $w(t)$,
respectively, and where $G_u(\lambda)$, $G_d(\lambda)$,
$G_f(\lambda)$   and $G_w(\lambda)$ are the \emph{transfer-function matrices} (TFMs) from the respective
inputs to outputs. For simplicity, we will assume that all these TFMs are \emph{proper} (i.e., finite for $\lambda = \infty$). Input-output models with additive faults of the form (\ref{systemw}) are useful in formulating various fault diagnosis problems, in deriving general solvability conditions and in describing conceptual synthesis procedures. However, these models are generally not suited for numerical computations, due to the  potentially high sensitivity of polynomial-based model representations.

For computational purposes, instead of the input-output model (\ref{systemw}) with the compound TFM $[\, G_u(\lambda) \; G_d(\lambda) \; G_f(\lambda) \; G_w(\lambda) \,]$, an equivalent state-space model is used having the form
\be\label{ssystem1} {\arraycolsep=1mm\begin{array}{rcl} E\lambda x(t) &=& Ax(t) + B_u u(t) + B_d d(t) +  \\ && \phantom{Ax(t) + \;}B_f f(t) + B_w w(t) \, ,  \\
y(t) &=&Cx(t) + D_u u(t) + D_d d(t) +  \\ &&\phantom{Cx(t) + \;}D_f f(t) + D_w w(t)  \, ,
\end{array}} \ee
 with the $n$-dimensional state vector $x(t)$, where $\lambda
x(t) := \dot{x}(t)$ or $\lambda x(t) := x(t+1)$ depending on the
type of the system, continuous- or discrete-time, respectively.  The matrix $E$ is generally invertible and is frequently taken as $E = I_n$. Plant models of the form (\ref{ssystem1})  often arise from the linearization of nonlinear dynamic plant models in specific operation points and for  fixed values of plant parameters. The noise inputs frequently  account  for the effects of uncertainties (e.g., inherent  variabilities in operating points and parameters).

\emph{Notation: }To indicate the input-output equivalence of the models in (\ref{systemw}) and (\ref{ssystem1}), we use the notation
{\small
\[ [\, G_u(\lambda) \; G_d(\lambda) \; G_f(\lambda) \; G_w(\lambda) \,] = {\arraycolsep=0.5mm\ba{c|cccc} A-\lambda E & B_u & B_d & B_f & B_w \\  \hline
C & D_u & D_d & D_f & D_w \ea}. \]
}

\section{Residual Generation}

A nonzero fault signal, $f \neq 0$, in (\ref{systemw}) signifies a deviation from the normal behavior of the plant
due to an unexpected event (e.g., physical component failure or
supply breakdown). Generally, the occurrence of a fault
must be detected as early as possible to prevent further degradation
of the plant behavior. The {fault diagnosis} techniques
are used to perform the {detection} of occurrence of faults (\emph{fault detection}), the
localization of detected faults (\emph{fault isolation}), the reconstruction of the fault signal (\emph{fault estimation}) and the classification of the
detected faults and determination of their characteristics (\emph{fault identification}). In a specific practical application, the term
\emph{fault detection and diagnosis} (FDD) may include, besides fault detection,  also further aspects such as fault isolation, fault estimation, or fault identification (not discussed in this article).

\begin{figure}[thpb]
\begin{center}
\includegraphics[height=4.5cm]{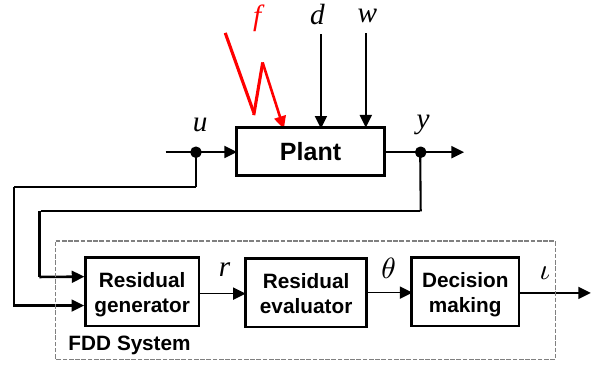}
\caption{Basic fault diagnosis setup.}  
\label{Fig_FDDSystem}                                 
\end{center}                                 
\end{figure}

A FDD system is a device (usually based on a collection of real-time processing algorithms)
suitably set up to fulfill the above tasks. The minimal functionality of any FDD system is illustrated in Fig.~\ref{Fig_FDDSystem}.

The main component of any FDD system is the \emph{residual
generator} (or \emph{fault detection filter}),
which produces residual signals grouped in a $q$-dimensional vector $r$ by processing the available
measurements $y$ and the known values of control inputs $u$. The role of the residual signals
is to indicate the presence  or absence of faults, and therefore the residual $r$ must be equal (or
close) to zero in the absence of faults and significantly different
from zero after a fault occurs.
For decision-making, suitable measures of the residual magnitudes are generated in a scalar or vector $\theta$  (e.g., $\theta = \|r\|$) , which is then used to produce the corresponding  decision variable or vector $\iota$ (e.g., $\iota= 1$ if $\theta > \tau$ for a detected fault and $\iota= 0$ if $\theta \leq \tau$  for the lack of faults, where $\tau$ is a given detection threshold). If $r(t)$ is a structured vector with, say $n_b$ components $r^{(i)}(t)$, $i = 1, \ldots, n_b$, then $\theta$ and $\iota$ are $n_b$-dimensional vectors, with $\theta_i$ representing a measure of the magnitude of the $i$-th residual component (e.g., $\theta_i = \|r^{(i)}\|$) and the binary signature $\iota_i = 1$ or $\iota_i = 0$ corresponding to a fired (i.e, $\theta_i > \tau$) or not fired (i.e., $\theta_i \leq \tau$)  component $r^{(i)}(t)$, respectively.

A linear residual generator employed in the FDD system in Fig.~\ref{Fig_FDDSystem} has the input-output form
 \be\label{detec}
{\mathbf{r}}(\lambda) = Q(\lambda)\ba{c}
{\mathbf{y}}(\lambda)\\{\mathbf{u}}(\lambda)\ea ,\ee where
$Q(\lambda)$ is the TFM of the filter. For a physically
realizable filter,  $Q(\lambda)$ must be \emph{stable} (i.e., proper and only with
poles having negative real parts for a continuous-time system
or magnitudes less than one for a discrete-time system). The
(dynamic) \textit{order} of $Q(\lambda)$ (also known as
\textit{McMillan degree}) is the dimension of the state vector
of a minimal state-space realization of $Q(\lambda)$. The
dimension $q$ of the residual vector $r(t)$ depends on the
fault diagnosis problem to be solved. The form (\ref{detec}) of the fault detection filter is called the \emph{implementation form} and is the basis of the real-time implementation of the fault detection filter.

The residual signal $r(t)$ in (\ref{detec}) generally depends
via the system outputs $y(t)$ of all system inputs $u(t)$,
$d(t)$, $f(t)$ and $w(t)$. The \emph{internal form} of the filter is obtained by replacing in (\ref{detec})
${\mathbf{y}}(\lambda)$ by its expression in (\ref{systemw}), and is given by
\begin{align} \label{resys}
{\mathbf{r}}(\lambda) = & \;\;
R_u(\lambda){\mathbf{u}}(\lambda) +
R_d(\lambda){\mathbf{d}}(\lambda) + \nonumber \\ & \;\;
R_f(\lambda){\mathbf{f}}(\lambda) + R_w(\lambda){\mathbf{w}}(\lambda)
,
\end{align}
where
\begin{align} [\, R_u(\lambda) \mid  R_d(\lambda) \mid  R_f(\lambda) \mid  R_w(\lambda)\,] := \qquad\qquad\nonumber \\  \;\; 
Q(\lambda)  \ba{c|c|c|c} G_u(\lambda) & G_d(\lambda) & G_f(\lambda)  & G_w(\lambda) \\
         I_{m_u} & 0 & 0 & 0 \ea \label{resys1}.
\end{align}
For a successfully designed filter $Q(\lambda)$, all TFMs in the
corresponding internal form (\ref{resys}) are  stable, and  additionally achieve  specific fault detection requirements.

The basic functionality  of a well-designed fault detection filter is to ensure the lack of false alarms, in the case when no faults occurred, and the lack of missed detection of faults, in the case of occurrence of a fault. The first requirement is fulfilled provided the residual signal $r(t)$ is zero, if there are no faults (i.e., $f(t) = 0$) and no noise (i.e., $w(t) = 0$), and in the presence of arbitrary control and disturbance inputs. If noise is present, then, in the case of absence of faults, the signal norm $\|r\|$ must be sufficiently small for all possible control, disturbance and noise inputs (i.e., $\theta \leq \tau_a$ for a sufficiently small alarm threshold $\tau_a$). The requirement on the lack of missed detections is fulfilled provided $\|r\|$ is sufficiently large for any fault of sufficiently large amplitude for all possible control, disturbance and noise inputs (i.e., $\theta > \tau_d$ for a certain detection threshold $\tau_d \geq \tau_a$).

The  requirement on the lack of false alarms can be transcribed in concrete requirements on the TFMs in the internal form.
It is always possible to completely decouple the control input $u(t)$ from the residual $r(t)$ (i.e., to achieve  $R_u(\lambda) = 0$), and we impose a similar condition on the disturbance input $d(t)$ by requiring $R_d(\lambda) = 0$. To minimize the effect of the noise input $w(t)$ on the residual, we always aim to simultaneously minimize the norm of $R_w(\lambda)$. The  distinction between the unknown inputs $d(t)$ and $w(t)$ lies solely in the way these signals are treated when solving the residual generator synthesis problem. For example, if the decoupling of all components of $d(t)$ is not possible, then some components of $d(t)$ can be redefined as additional noise inputs.
The requirement on the lack of missed fault detection leads to additional requirements imposed on $R_f(\lambda)$ (see next section).
%

\section{Fault Diagnosis Problems}

In the literature dedicated to the solution of fault diagnosis problems we can observe a diversity of formulations of the basic fault diagnosis problems. The differences in problem formulations are partly due to the employed particular system theoretical frameworks and partly because of focusing on particular classes of solution methods. Therefore, it is important to formulate a (canonical) set of basic fault diagnosis problems which cover most practical applications.  In this endeavour, (at least) two aspects have to be considered.

A first aspect is related to addressing the effects of uncertainties when solving fault diagnosis problems. In this context, two categories of problems may arise. The exact synthesis problems have as goal the determination of fault detection filters which fulfill strict (algebraic) existence conditions. In these problems the  effects of the (ubiquitous) noise is fully neglected and the effects of (unknown) disturbances are exactly decoupled. The (more practice relevant) approximate synthesis problems solve essentially the same type of problems, with the additional goal of attenuating, as much as possible, the effects of the inherent noise. A meaningful requirement  in formulating fault diagnosis problems is to ensure that the formulations
of approximate problems include the formulations of exact problems if the effect of noise can be neglected. The expected consequence is that methods to solve approximate problems can be employed to solve exact problems too, while the solutions of exact problems can be also considered as candidate solutions of the approximate problems.

Another important aspect is to formulate the fault diagnosis problems independently of any particular solution method intended to be used to solve these problems. It is often the case that the applicability of a specific solution method involves additional (technical) conditions, which however are not necessary conditions for the solvability of the fault diagnosis problem. Therefore, to avoid such additional constraints, it is important to have problem formulations for which rigourous (ideally, necessary and sufficient) existence conditions of the solution exist, which guarantee solvability  in the most general setting.

In all fault diagnosis problems formulated in what follows, we require that by a suitable choice of a stable fault detection filter $Q(\lambda)$, we achieve that  the residual signal $r(t)$ is fully decoupled from the control input $u(t)$ and disturbance input $d(t)$. Thus, the following \emph{decoupling conditions} must be generally fulfilled:
\be\label{ens} \begin{array}{ll}
  (i) & R_u(\lambda) = 0 ,\\
  (ii) & R_d(\lambda) = 0 .
\end{array}
\ee

For each fault diagnosis problem specific requirements have to be fulfilled, which are formulated as additional conditions in what follows. For the formulated problems  we also give the existence conditions of the solutions of these problems in terms of some rank conditions.

\subsection*{Exact fault detection problem  -- EFDP} \label{sec:EFDP}
The basic additional requirement is simply to achieve by a suitable choice of a fault detection filter $Q(\lambda)$ that in the absence of noise input (i.e., $w(t) \equiv 0$), the residual  $r(t)$ is influenced by all fault components $f_j(t)$, $j = 1, \ldots, m_f$. Let $R_{f_j}(\lambda)$ denote the $j$-th column of $R_{f}(\lambda)$. This requirement can be expressed as the following \emph{detection condition} to be fulfilled for all faults:
\be\label{efdp}
 \hspace*{-2mm} (iii) \; R_{f_j}(\lambda) \not = 0, j = 1, \ldots, m_f \;\; \textrm{with} \; R_f(\lambda) \; \textrm{stable.}
\ee

The solvability conditions of the EFDP are simple rank conditions involving only the TFMs from the disturbances and faults:
\begin{theorem}\label{T1}
For the system (\ref{systemw}) with $w(t) \equiv 0$  the EFDP is solvable if and only
if
\be\label{fdcond} \hspace*{-2mm}\rank  [\, G_d(\lambda) \; G_{f_j}(\lambda)\, ] > \rank  G_d(\lambda), \; j = 1, \ldots , m_f ,
\ee
where $G_{f_j}(\lambda)$ denotes the $j$-th column of $G_{f}(\lambda)$.
\end{theorem}
Here, $\rank G(\lambda)$ denotes the normal rank of the rational TFM $G(\lambda)$, representing the maximal rank of the complex matrix $G(\lambda)$ over all values of $\lambda \in \mathds{C}$ such that $G(\lambda)$ has finite norm.

The conditions (\ref{fdcond}) define the \emph{complete fault detectability} property of the system (\ref{systemw}). These conditions are generically fulfilled if
$p > m_d$, that is, when there are more measurements than disturbance inputs.
The importance of solving EFDPs for fault diagnosis primarily lies in the fact that the solution of the (more involved) fault isolation problems can be addressed by successively solving several appropriately formulated EFDPs.

%

\subsection*{Approximate fault detection problem  -- AFDP}\label{sec:AFDP}
The effects of the noise input $w(t)$ can usually not be fully decoupled from the residual $r(t)$. In this case, the basic requirements for the choice of $Q(\lambda)$  can be expressed to achieve that the residual  $r(t)$ is  influenced by all fault components $f_j(t)$ and the influence of the noise signal $w(t)$ is negligible. Thus, the following two additional conditions have to be fulfilled:
\be\label{afdp} \hspace*{-2mm}{\arraycolsep=0.5mm \begin{array}{ll}
  (iii) & R_{f_j}(\lambda) \not = 0, j = 1, \ldots, m_f \; \textrm{with} \; R_f(\lambda) \; \textrm{stable,} \\
  (iv) & R_w(\lambda) \approx 0, \;\; \textrm{with} \; R_w(\lambda) \; \textrm{stable.}
\end{array} }
\ee
Here, $(iii)$ is the detection condition of all faults employed also in the EFDP, while $(iv)$ is the  \emph{attenuation condition} for the noise input. The condition $R_w(\lambda) \approx 0$ expresses the requirement  that the transfer gain $\|R_w(\lambda)\|$ (measured by any suitable norm) can be made arbitrarily small.

The solvability conditions of the formulated AFDP are simply those of the EFDP:
\begin{theorem}\label{T2}
For the system (\ref{systemw}) the AFDP is solvable if and only if the EFDP is solvable.
\end{theorem}
On the basis of this result, any properly scaled solution of the EFDP can potentially serve as a solution of the AFDP as well.
At signal level, a bounded noise input $w(t)$ such that $\|w \| \leq \delta_w$, can have a maximum contribution $\|R_w(\lambda)\|\delta_w$ in the residual $r(t)$, which automatically determines the minimum size $\delta_{f,min}$ of detectable faults. For example, $\delta_{f,min}$ can be computed as $\delta_{f,min} = \frac{\delta_w}{\eta}$, where
\be\label{nfgap} \eta := \frac{\displaystyle\min_{1\leq j \leq m_f} \|R_{f_j}(\lambda)\|}{\|R_w(\lambda)\|} \ee
is the fault-to-noise gap.
The resulting value of $\delta_{f,min}$ can be used to assess the ``practical'' usefulness of any solution, and the maximization of the gap $\eta$ is always a meaningful goal for the synthesis of fault detection filters.

\subsection*{Exact fault detection and isolation problem -- EFDIP} \label{sec:EWFDIP}
For the isolation of faults, we employ residual generator filters formed by stacking a bank of $n_b$ filters of the form
\be\label{ri_fdip}
{\mathbf{r}}^{(i)}(\lambda) = Q^{(i)}(\lambda)\ba{c}
{\mathbf{y}}(\lambda)\\{\mathbf{u}}(\lambda)\ea \, ,\ee
where the $i$-th filter $Q^{(i)}(\lambda)$ generates the corresponding $i$-th residual component $r^{(i)}(t)$ (scalar or vector). This leads to the following structured residual vector $r(t)$ and block-structured filter $Q(\lambda)$
\be\label{qbank}
r(t) = \ba{c} r^{(1)}(t)\\ \vdots \\ r^{(n_b)}(t) \ea , \;
Q(\lambda) = \ba{c} Q^{(1)}(\lambda)\\ \vdots \\ Q^{(n_b)}(\lambda) \ea  \, .\ee
The resulting $R_f(\lambda)$ is an $n_b\times m_f$ block-structured TFM of the form
\be\label{Rf_struct} R_f(\lambda) = \ba{ccc} R^{(1)}_{f_1}(\lambda)& \cdots &R^{(1)}_{f_{m_f}}(\lambda) \\
\vdots & \ddots & \vdots \\
 R^{(n_b)}_{f_1}(\lambda)& \cdots &R^{(n_b)}_{f_{m_f}}(\lambda) \ea \, , \ee
 where the $(i,j)$-th block
of $R_f(\lambda)$ is defined as \[ R^{(i)}_{f_j}(\lambda) := Q^{(i)}(\lambda) \ba{c} G_{f_j}(\lambda) \\ 0 \ea \]  and   describes how the $j$-th fault $f_j$ influences the $i$-th residual component $r^{(i)}(t)$.

We associate to the block-structured $R_f(\lambda)$ in (\ref{Rf_struct}) the $n_b\times m_f$ binary
\emph{structure
matrix} $S_{R_f}$, whose $(i,j)$-th element is defined as
\be\label{structure_matrix}
{\arraycolsep=1mm\begin{array}{llrll} S_{R_f}(i,j) &=& 1 & \text{ if } & R^{(i)}_{f_j}(\lambda) \not=0 \; ,\\
S_{R_f}(i,j) &=& 0 & \text{ if } & R^{(i)}_{f_j}(\lambda) =0 \, .
\end{array}} \ee
If $S_{R_f}(i,j) = 1$
then we say that the residual component $r^{(i)}$ is sensitive to the $j$-th fault $f_j$, while if $S_{R_f}(i,j) = 0$ then the $j$-th fault $f_j$  is decoupled from $r^{(i)}$. The $m_f$ columns of $S_{R_f}$ are called \emph{fault signatures}. Since each nonzero column of $S_{R_f}$ is associated with the corresponding fault input, fault isolation can be performed by comparing the resulting binary decision vector $\iota$ in Fig.~\ref{Fig_FDDSystem} (i.e., the signatures of fired or not fired residual components) with the fault signatures coded in the columns of $S_{R_f}$.

%
%
%

In the absence of noise input (i.e., $w(t) \equiv 0$) and for a given $n_b\times m_f$ structure matrix $S$, the EFDIP  requires the determination of   a stable  filter $Q(\lambda)$ of the form (\ref{qbank}) such that for the corresponding block-structured $R_f(\lambda)$ in (\ref{Rf_struct}), the following condition is additionally fulfilled:
\[
  (iii) \;\; S_{R_f} = S , \;\; \textrm{with} \; R_f(\lambda) \; \textrm{stable.}
\]

The solution of the EFDIP can be addressed by solving $n_b$ suitably formulated EFDPs. The $i$-th EFDP arises by reformulating the $i$-th EFDIP for determining the $i$-th  filter $Q^{(i)}(\lambda)$ in (\ref{ri_fdip}) for a structure matrix which is the $i$-th row of $S$. This can be accomplished by redefining the fault components corresponding to zero entries in the $i$-th row of $S$ as additional disturbance inputs to be decoupled in the $i$-th residual component $r^{(i)}(t)$.
%
%
%
Let $\widehat G_d^{(i)}(\lambda)$ be the TFM formed from the columns of $G_f(\lambda)$ for which $S_{ij} = 0$. We have the following solvability conditions for the EFDIP, which simply express the solvability of the $n_b$ EFDPs formulated for the $n_b$ rows of of $S$:

\begin{theorem}\label{T4}
For the system (\ref{systemw}) with $w(t) \equiv 0$   and a given $n_b\times m_f$ structure matrix $S$,  the EFDIP is solvable  if and only if for $i = 1, \ldots, n_b$
\be\label{fdcondri} \textrm{rank}\, [\, G_d(\lambda) \; \widehat G_d^{(i)}(\lambda)\;
G_{f_j}(\lambda)\, ] > \textrm{rank}\, [\, G_d(\lambda) \; \widehat G_d^{(i)}(\lambda)\, ] \ee
for all $j$ such that $S_{ij} \not = 0$.
\end{theorem}
The conditions (\ref{fdcondri}) define the \emph{$S$ fault isolability} property of the system (\ref{systemw}). If $S$ has at most $k$ zero entries in each row, then these conditions are generically fulfilled if
$p > m_d+k$, that is, when the number of measurements exceeds the number of  disturbance inputs with at least $k$.
An important case is when $S = I_{m_f}$  (i.e., diagonal), in which case $k = m_f-1$. If the conditions (\ref{fdcondri}) are fulfilled, then the system (\ref{systemw}) is called \emph{strongly isolable}. For a strongly isolable system we have the following solvability conditions:
\begin{theorem}\label{T-strongEFDIP}
For the system (\ref{systemw}) with $w \equiv 0$ and $S = I_{m_f}$, the EFDIP is solvable  if and only if \be\label{fdistrong} \rank \, [\, G_d(\lambda) \;
G_{f}(\lambda)\, ] = \rank  G_d(\lambda) + m_f  \, .\ee
\end{theorem}
Generically, the condition (\ref{fdistrong}) is fulfilled if
$p\geq m_f+m_d$, which implies that the system must have at least as many measurements as the total number of disturbance and fault inputs. The importance of strong isolability is that it allows to isolate the occurrence of an arbitrary number (up to $m_f$) simultaneous faults.

\subsection*{Approximate fault detection and isolation problem  -- AFDIP}

Let $S$ be a desired $n_b\times m_f$ structure matrix targeted to be achieved by using a structured fault detection filter $Q(\lambda)$ with $n_b$ row blocks as in (\ref{qbank}). The $n_b\times m_f$ block-structured TFM $R_f(\lambda)$, corresponding to $Q(\lambda)$, is defined in (\ref{Rf_struct}). $R_f(\lambda)$
can be additively decomposed as $R_f(\lambda) = \widetilde  R_f(\lambda) + \overline R_f(\lambda)$, where  $\widetilde  R_f(\lambda)$ and $\overline R_f(\lambda)$ have the same block structure as $R_f(\lambda)$ and have their $(i,j)$-th blocks defined as
\[ \widetilde  R^{(i)}_{f_j}(\lambda) := S_{ij}R^{(i)}_{f_j}(\lambda), \quad \overline R^{(i)}_{f_j}(\lambda) := (1-S_{ij})R^{(i)}_{f_j}(\lambda) \, .\]
To address the approximate fault detection and isolation problem, we will target to enforce for
the part $\widetilde R_f(\lambda)$ of $R_f(\lambda)$ the desired structure matrix $S$, while the part $\overline R_f(\lambda)$ must be (ideally) negligible.
The \emph{soft approximate fault detection and isolation problem} (\emph{soft} AFDIP) can be formulated as follows. For a given $n_b\times m_f$ structure matrix $S$, determine a stable and proper filter $Q(\lambda)$  in the form (\ref{qbank}) such that the following conditions are additionally fulfilled:
\be\label{afdip-sr} \hspace*{-7mm}{\arraycolsep=1mm\begin{array}{ll}
  (iii) & S_{\widetilde R_f} = S, \; \overline R_f(\lambda) \approx 0 , \; \text{with} \; R_f(\lambda) \; \text{stable,} \;  \\
    (iv) & R_w(\lambda) \approx 0, \; \text{with} \; R_w(\lambda) \; \text{stable.}
\end{array}}
\ee

The following (somewhat surprising) result states that the solvability condition of the AFDIP is precisely the solvability of the EFDP.

\begin{theorem}\label{T-AFDIP}
For the system (\ref{systemw}) and a given structure matrix $S$ without zero columns, the \emph{soft} AFDIP is solvable if and only if the EFDP is solvable.
\end{theorem}

The solvability of the EFDIP is clearly a sufficient condition for the solvability of the \emph{soft} AFDIP, but is not, in general, also a necessary condition, unless we impose in the formulation of the AFDIP the stronger condition $\overline R_f(\lambda) = 0$ (instead  $\overline R_f(\lambda) \approx 0$). This is equivalent to require $S_{R_f} = S$. Therefore, we can alternatively formulate the \emph{strict} AFDIP  to fulfill the conditions:
\be\label{afdip-sr1} \hspace*{-7mm}{\arraycolsep=1mm\begin{array}{ll}
  (iii)' & S_{R_f} = S, \; \text{with} \; R_f(\lambda) \; \text{stable,} \;  \\
    (iv)' & R_w(\lambda) \approx 0, \; \text{with} \; R_w(\lambda) \; \text{stable.}
\end{array}}
\ee
In this case  we have the (expected) result:
\begin{theorem}\label{T-AFDIPE}
For the system (\ref{systemw}) and a given structure matrix $S$, the \emph{strict} AFDIP is solvable with $S_{R_f} = S$ if and only if the EFDIP is solvable.
\end{theorem}

\subsection*{Exact model-matching problem -- EMMP}

Let $M_r(\lambda)$ be a given $q\times m_f$ TFM of a stable reference model  specifying the desired input-output behavior from the faults to residuals as
\[ {\mathbf{r}}(\lambda) = M_r(\lambda) {\mathbf{f}}(\lambda). \]
 Thus, we want to achieve by a suitable choice of a stable $Q(\lambda)$  satisfying $(i)$ and $(ii)$ in (\ref{ens}), that we have additionally $R_f(\lambda) = M_r(\lambda)$. For example, a typical choice for $M_r(\lambda)$ is an $m_f \times m_f$  diagonal and invertible TFM, which ensures that each residual $r_i(t)$ is influenced only by the fault $f_i(t)$. This would allow the isolation of arbitrary combinations of up to $m_f$ simultaneous faults.  The choice $M_r(\lambda) = I_{m_f}$ targets the solution of an  \emph{exact fault estimation problem} (EFEP).

To determine $Q(\lambda)$, we have to solve the linear rational equation (\ref{resys1}), with the settings $R_u(\lambda) = 0$, $R_d(\lambda) = 0$, and $R_f(\lambda) = M_r(\lambda)$ ($R_w(\lambda)$ and $G_w(\lambda)$ are assumed empty matrices). The choice of $M_r(\lambda)$ may lead to a solution $Q(\lambda)$ which is not proper or is unstable or has both these undesirable properties. Therefore, besides determining $Q(\lambda)$, we also consider the determination of a suitable updating factor $M(\lambda)$ of $M_r(\lambda)$ to ensure the stability of the solution $Q(\lambda)$ for $R_f(\lambda) = M(\lambda) M_r(\lambda)$. Obviously, $M(\lambda)$ must be chosen a stable and invertible TFM. Additionally,  by choosing $M(\lambda)$ diagonal, the zero and nonzero entries of $M_r(\lambda)$ can be also preserved in $R_f(\lambda)$ (i.e., to cope with the formulation of the EFDIP).

To address the above aspect, the EMMP can be formulated to also include the selection of a diagonal, stable and invertible TFM $M(\lambda)$ such that the following condition is additionally fulfilled:
\be\label{emmp}
  (iii) \;\; R_f(\lambda) = M(\lambda)M_r(\lambda)\, .\ee

The conditions (\ref{ens}) and (\ref{emmp}) represent a linear system of rational equations of the form
\be\label{esfdip-full} Q(\lambda)G_e(\lambda) = [\,  0 \;\; 0 \;\; M(\lambda)M_r(\lambda) \,]\, , \ee
where
\be\label{Ge} G_e(\lambda) := \ba{ccc} G_u(\lambda) & G_d(\lambda) & G_f(\lambda) \\ I_{m_u} & 0 & 0 \ea . \ee
Therefore, the solvability condition of the EMMP follows from the standard solvability condition of systems of linear equations:
\begin{theorem}\label{T-EMMP}
\index{model-matching problem!a@exact (EMMP)!solvability|ii}
For the system (\ref{systemw}) with $w \equiv 0$   and a given $M_r(\lambda)$, the EMMP is solvable if and only if the following condition is fulfilled
\be\label{fdimmcond} \hspace*{-1mm}
\rank\, [\, G_f(\lambda)\; G_d(\lambda)\, ] =
\rank \, {\arraycolsep=0.5mm\ba{cc} G_f(\lambda) & G_d(\lambda)\\ M_r(\lambda) & 0 \ea} \, .\\
\ee
\end{theorem}

When $M_r(\lambda)$ has full column rank $m_f$, the solvability condition (\ref{fdimmcond}) of the EMMP  reduces to the strong isolability condition (\ref{fdistrong}) of Theorem~\ref{T-strongEFDIP}.

The solvability conditions become more involved if we strive for a stable solution $Q(\lambda)$ for a given reference model $M_r(\lambda)$ without allowing its updating. For example, this is the case when solving the EFEP for $M_r(\lambda) = I_{m_f}$, in which case, we have the following result.

\begin{theorem}\label{T-FEP}
For the system (\ref{systemw}) with $w \equiv 0$, the EFEP is solvable if and only if  the system is strongly fault isolable and $G_f(\lambda)$ is minimum phase.
\end{theorem}

\subsection*{Approximate model-matching problem  -- AMMP}

Similarly to the formulation of the EMMP, we include the determination of an updating factor of the reference model in the formulation of the AMMP. Specifically, for a given stable TFM $M_r(\lambda)$, it is required to determine a stable  filter $Q(\lambda)$ and a stable and invertible diagonal $M(\lambda)$ such that the following conditions are additionally fulfilled:
\be\label{afdip-mm} \hspace*{-2mm}{\arraycolsep = 0.5mm\begin{array}{ll}
  (iii) & R_f(\lambda) \approx M(\lambda)M_r(\lambda), \;\; \textrm{with} \;\; R_f(\lambda) \;\; \textrm{stable};\\
  (iv) & R_w(\lambda) \approx 0, \;\; \textrm{with} \;\; R_w(\lambda) \;\; \textrm{stable.}
\end{array}}
\ee

Necessary and sufficient conditions for the solution of the AMMP are not known. However, a straightforward  sufficient condition for the solvability of the AMMP is simply the solvability of the EMMP. Moreover, any solution of an exact problem (e.g., EFDP, EFDIP or EMMP) generates a meaningful reference model $M_r(\lambda) := R_f(\lambda)$, which can serve to improve the noise attenuation performance by solving an an appropriately formulated  AMMP.

\section{Synthesis of Fault Detection Filters} \label{ch:fdsynth}

The recently developed computational procedures for the synthesis of fault detection filters \citep{Varg17} share several computational paradigms, which are instrumental in developing generally applicable, numerically reliable and computationally efficient synthesis methods.
In what follows we shortly review these paradigms and discuss their roles in the synthesis procedures.

\subsection*{Nullspace-based synthesis}
An important synthesis paradigm is the use of the nullspace method as a first synthesis step to ensure the fulfillment of the decoupling conditions  $R_u(\lambda) = 0$ and $R_d(\lambda) = 0$ in (\ref{ens}). This can be done by choosing $Q(\lambda)$ of the form
\be\label{parQ} Q(\lambda) = \overline Q_1(\lambda) Q_1(\lambda) , \ee%
where the factor $Q_1(\lambda)$ is a left annihilator of
\be\label{null} G(\lambda) := \ba{cc} G_u(\lambda) & G_d(\lambda) \\ I_{m_u} & 0 \ea \, .\ee
For any $Q(\lambda)$ of the form (\ref{parQ}),
 we have
\[ [\,R_u(\lambda) \; R_d(\lambda)\,] = Q(\lambda)G(\lambda)= 0 \, .\]
Assume  $r_d < p$ is the normal rank of $G_d(\lambda)$. Using standard linear algebra results, there exists a maximal full row rank left annihilator  $N_l(\lambda)$ of size $(p-r_d)\times (p+m_u)$ such that $N_l(\lambda)G(\lambda) = 0$. Any such an $N_l(\lambda)$ represents a \emph{basis} of the left nullspace of the rational matrix $G(\lambda)$. With the choice $Q_1(\lambda) = N_l(\lambda)$, the expression (\ref{parQ}) provides a parametrization of solutions of \emph{all} fault diagnosis problems formulated in the previous section. For example, $N_l(\lambda)$ can be chosen in the product form
\[
N_l(\lambda) = N_{l,d}(\lambda) [\, I_p \; -\!G_u(\lambda)\,] \]
where $N_{l,d}(\lambda)$ is a left nullspace  basis of $G_d(\lambda)$ satisfying $N_{l,d}(\lambda)G_d(\lambda) = 0$.
For the synthesis of fault detection filters an important aspect is to use proper left nullspace bases of least dynamical orders. Such a basis $N_l(\lambda)$ has no finite or infinite zeros (i.e., $\rank N_l(\lambda) = p-r_d$ for all $\lambda \in \mathds{C}$) and can be chosen having arbitrary poles (e.g., stable).

The form (\ref{parQ}) of the filter allows to reformulate all synthesis problems as simpler problems (without control and disturbance inputs), which allow to easily check the solvability conditions.
With (\ref{parQ}), the fault detection filter in (\ref{detec}) can be rewritten in the alternative form
 \be\label{detec1}
{\mathbf{r}}(\lambda) = \overline Q_1(\lambda)Q_1(\lambda)\ba{c}
{\mathbf{y}}(\lambda)\\{\mathbf{u}}(\lambda)\ea = \overline Q_1(\lambda) \overline{\mathbf{y}}(\lambda) \;, \ee
where
\begin{align} \label{systemfw} \overline{\mathbf{y}}(\lambda) &:= Q_1(\lambda)\ba{c}
{\mathbf{y}}(\lambda)\\{\mathbf{u}}(\lambda)\ea \nonumber \\  &= \overline G_f(\lambda){\mathbf{f}}(\lambda) +
\overline G_w(\lambda){\mathbf{w}}(\lambda)  \,,
\end{align}
with
\be\label{Gfw} [\, \overline G_f(\lambda) \; \overline G_w(\lambda) \,] := Q_1(\lambda)
{\arraycolsep=1mm\ba{cc} G_f(\lambda) & G_w(\lambda) \\ 0 & 0 \ea }\, .\ee
With this first preprocessing step, we reduced the original problems formulated for the system (\ref{systemw}) to simpler ones, which can be formulated for the reduced system (\ref{systemfw}) (without control and disturbance inputs),  for which we have to determine the TFM $\overline Q_1(\lambda)$ of the simpler fault detection filter (\ref{detec1}).

At this stage we can assume that both  $Q_1(\lambda)$ and the  TFMs of the reduced system (\ref{systemfw}) are proper and even stable. This can be always achieved by replacing any basis $N_l(\lambda)$, with a stable basis $Q_1(\lambda) = M(\lambda)N_l(\lambda)$, where $M(\lambda)$ is an invertible, stable TFM (e.g.,  of least McMillan degree), such that $M(\lambda)[\,N_l(\lambda)\;\overline G_f(\lambda)\; \overline G_w(\lambda)\,]$ is stable. Such an $M(\lambda)$ can be determined as the (minimum-degree) denominator of a stable left coprime factorization of $[\,N_l(\lambda)\;\overline G_f(\lambda)\; \overline G_w(\lambda)\,]$.

All synthesis problems can be equivalently reformulated to determine a filter $\overline Q_1(\lambda)$ for the reduced system (\ref{systemfw}). As an example, we give the simpler conditions for the solvability of the EFDP (and also of the AFDP and \emph{soft} AFDIP).

\begin{corollary}  For the system (\ref{systemw}) with $w \equiv 0$, let $Q_1(\lambda)$ be a rational basis of the left nullspace of $G(\lambda)$ in (\ref{null}), and let (\ref{systemfw}) be the corresponding reduced system with $w \equiv 0$.  Then, the EFDP is solvable if and only if
\be\label{fdetec_red}
\overline G_{f_j}(\lambda) \not = 0, \quad j = 1, \ldots m_f \, .\ee
\end{corollary}
Similar simpler conditions can be derived for the solvability of the EFDIP and \emph{strict} AFDIP.

Particularly simple becomes the solvability condition of the EFDIP with strong isolability  requirement (and also of the EMMP with invertible $M_r(\lambda)$) as the left invertibility of $\overline G_f(\lambda)$.
\begin{corollary} For the system (\ref{systemw})  with $w \equiv 0$ and $S = I_{m_f}$, the EFDIP is solvable if and only if for the reduced system (\ref{systemfw}) with $w \equiv 0$  we have
\[ \rank \overline G_f(\lambda) = m_f \;.\]
\end{corollary}

\subsection*{Filter updating techniques}

In many synthesis procedures the TFM of the resulting filter $Q(\lambda)$ can be expressed in a factored form as
\be\label{qfact} Q(\lambda) = Q_K(\lambda) \cdots Q_2(\lambda)Q_1(\lambda) \, , \ee
where $Q_1(\lambda)$ is a left nullspace basis of $G(\lambda)$ in (\ref{null}) satisfying $Q_1(\lambda)G(\lambda) = 0$, and  $Q_1(\lambda)$, $Q_2(\lambda)Q_1(\lambda)$, $\ldots$, can be interpreted as partial syntheses addressing specific requirements. Since each partial synthesis may represent a valid fault detection filter, this approach has a high flexibility in using or combining different synthesis techniques.

 The determination of $Q(\lambda)$ in the factored form (\ref{qfact}) can be formulated as a $K$-step synthesis procedure based on successive updating of an initial filter $Q = Q_1(\lambda)$ and the nonzero terms of its corresponding  internal form
 \[ R(\lambda) := [\, R_f(\lambda) \; R_w(\lambda)\,] = Q_1(\lambda){\arraycolsep=0.5mm\ba{cc} G_f(\lambda) & G_w(\lambda) \\  0 & 0 \ea} \]
as follows:
for $k = 2, \ldots, K$, do

\emph{Step $k$:} Determine $Q_k(\lambda)$ using the current $Q(\lambda)$ and $R(\lambda)$ and  then perform the updating as
 \[ Q(\lambda) \leftarrow Q_k(\lambda)Q(\lambda), \quad R(\lambda) \leftarrow Q_k(\lambda)R(\lambda) .\]

The updating operations are efficiently performed using state-space description based formulas.
The state-space realizations of the TFMs $Q(\lambda)$ and $R(\lambda)$ in the implementation and internal forms can be jointly expressed in the generalized  state-space (or descriptor system) representation
\be\label{QR}
[\, Q(\lambda) \; R(\lambda) \,] = \ba{c|cc} A_Q-\lambda E_Q & B_Q & B_R \\ \hline
C_Q & D_Q & D_R \ea .\ee
The state-space realizations of $Q(\lambda)$ and $R(\lambda)$ share the matrices $A_Q$, $E_Q$ and $C_Q$.  The preservation of these forms at each updating step is the basis of the filter updating based synthesis paradigm. To allow operations with non-proper TFMs (i.e., with singular $E_Q$), the use of the descriptor system representation is instrumental for developing numerically reliable computational algorithms.

The main benefit of using explicit state-space based updating formulas is the possibility to ensure at each step the cancellation of a maximum number of poles and zeros between the factors. This allows to keep the final order of the filter $Q(\lambda)$ as low as possible. In this way, the updating-based synthesis approach naturally leads to so-called \emph{integrated computational algorithms}, with strongly coupled successive computational steps. Since the form of the realizations of $Q(\lambda)$ and $R(\lambda)$ in (\ref{QR}) are preserved, the stability is simultaneously achieved for both  $Q(\lambda)$ and $R(\lambda)$ at the final synthesis step. At the last step, the standard state-space realization of $Q(\lambda)$ is usually recovered, to facilitate its real-time implementation.

\subsection*{Least order synthesis}
The least order synthesis of FDI filters means to determine filters $Q(\lambda)$ with the least possible dynamical orders. This is a valuable synthesis goal which helps to reduce the computational burden related to the real-time implementation of the filters. The main tool to achieve least order synthesis is the solution of suitable \emph{minimal cover problems}. If $X_1(\lambda)$ and $X_2(\lambda)$ are rational matrices of the same column dimension,  then the \emph{left minimal cover problem} is to find $X(\lambda)$ and $Y(\lambda)$ such that
\be\label{lmincov} X(\lambda) = X_1(\lambda) + Y(\lambda) X_2(\lambda) , \ee
and the McMillan degree of $[\,X(\lambda) \; Y(\lambda) \,]$ is minimal. Two cases are relevant to solve synthesis problems of fault detection filters.

A typical second step in many synthesis procedures is to choose $Q_2(\lambda)$ such that
the product $Q_2(\lambda)Q(\lambda)$ has least dynamical order and, simultaneously, a certain \emph{admissibility} condition is fulfilled (usually involving the nonzero TFMs $R_f(\lambda)$ and $R_w(\lambda)$). Typically, after the first step, $Q(\lambda) := Q_1(\lambda)$, $R_f(\lambda) := \overline G_f(\lambda)$ and $R_w(\lambda) := \overline G_w(\lambda)$, and $Q_1(\lambda)$ is a left nullspace basis of $G(\lambda)$ in (\ref{null}).
The admissible choices of $Q_2(\lambda)$ depend on the subsequent steps of the employed particular synthesis procedure and express conditions related to fault detectability or some full rank conditions. For example, for the solution of the EFDP, the admissibility conditions  are
\be\label{admEFDP} Q_2(\lambda)R_{f_j}(\lambda) \not = 0, \;\;j = 1, \ldots, m_f, \ee which  guarantee the detectability of all fault components. For the solution of the AFDP, additionally to (\ref{admEFDP}), the full row rank condition on $Q_2(\lambda)R_w(\lambda)$  has to be fulfilled as well. For the solution of the EMMP by using an inversion based approach, the admissibility condition is the invertibility of $Q_2(\lambda)R_f(\lambda)$, while for the solution of the AMMP the admissibility condition is the full row rank of $Q_2(\lambda)[\,R_f(\lambda)\; R_w(\lambda)\,]$. With the exception of (\ref{admEFDP}), the admissibility conditions are not necessary, and are used only for convenience, to simplify the subsequent computational steps.

%
The determination of candidate solutions $Q_2(\lambda)$ such that $Q_2(\lambda)Q(\lambda)$ has least order  can be done by solving left minimal cover problems of the form (\ref{lmincov}), where $X_1(\lambda)$ and $X_2(\lambda)$ represent disjoint subsets of basis vectors, such that: $Q(\lambda) = \left[\begin{smallmatrix} X_1(\lambda) \\ X_2(\lambda) \end{smallmatrix}\right]$, $Q_2(\lambda) = [\, I \;\; Y(\lambda)\,]$,  and $X(\lambda) = Q_2(\lambda)Q(\lambda)$ and $Y(\lambda)$ represent the solution of the left cover problem. A systematic search over increasing orders of candidate solutions can be performed and the search stops when the admissibility conditions are fulfilled.

The cover technique also allows to determine least order solutions of the EMMP, where $Q(\lambda)$ satisfies the rational linear equation (\ref{esfdip-full}).
The general solution of (\ref{esfdip-full}) can be expressed as
\[ Q(\lambda) = X_1(\lambda) + Y(\lambda) X_2(\lambda) ,\]
where  $X_1(\lambda)$ is a particular solution of (\ref{esfdip-full}) and $X_2(\lambda)$ is a left nullspace basis of $G_e(\lambda)$ in (\ref{Ge}). Therefore, a least order solution $Q(\lambda)$ can be computed by solving a left cover problem.

State-space representation based computational methods for the solution of minimum dynamic cover problems are described in \citep[Sections 7.5 and 10.4]{Varg17}, together with explicit updating formulas of the state-space realizations of  $Q(\lambda)$ and $R(\lambda)$.

\subsection*{Coprime factorization techniques}
A desired dynamics of the resulting final filters $Q(\lambda)$ and $R(\lambda)$ can be enforced by choosing a suitable invertible factor $M(\lambda)$, such that  $M(\lambda)[\, Q(\lambda) \; R(\lambda) \,]$ has desired poles. This can be achieved by computing a left coprime factorization
\[ [\, Q(\lambda) \; R(\lambda) \,] = M^{-1}(\lambda) [\, N_Q(\lambda) \; N_R(\lambda) \,] \]
with $M(\lambda)$ and $[\, N_Q(\lambda) \; N_R(\lambda) \,]$ coprime and having arbitrary stable poles, and then performing the updating operation
\[ [\, Q(\lambda) \; R(\lambda) \,] \leftarrow [\, N_Q(\lambda) \; N_R(\lambda) \,]  .  \]
The stabilization via a left coprime factorization is usually performed as the last step of the synthesis procedures. If the preservation of the zero-nonzero pattern of  $R_f(\lambda)$ is necessary (e.g., when solving fault isolation or model-matching problems), then $M(\lambda)$ can be  chosen block diagonal.

To illustrate the updating technique, we consider a realization of the denominator factor $M(\lambda)$ of the form
\[ M(\lambda) = \ba{c|c} A_Q+KC_Q-\lambda E_Q & K \\ \hline \\[-3mm]
C_Q & I \ea ,\]
where $K$ (the so-called output injection gain matrix) is determined by eigenvalue assignment techniques to ensure stability (i.e., to assign the poles of $M(\lambda)$ to lie in a suitable stable region).
Then, the numerator factors are obtained in forms compatible with $Q(\lambda)$ and $R(\lambda)$ in (\ref{QR}) as
\[ N_Q(\lambda) = \ba{c|c} A_Q+KC_Q-\lambda E_Q & B_Q+KD_Q \\ \hline \\[-3mm]
C_Q & D_Q \ea , \]
\[ N_R(\lambda) = \ba{c|c} A_Q\!+\!KC_Q-\lambda E_Q  & B_{R}+KD_{R} \\ \hline \\[-3mm]
C_Q  & D_{R}  \ea .\]

Instead of using an explicitly computed output injection matrix $K$, it is possible to employ alternative coprime factorization methods, based on recursive pole dislocation techniques \citep{Varg17d}, which produce directly the numerator factors $N_Q(\lambda)$ and $N_R(\lambda)$, and thus implicitly perform  the updating operations as well.

\subsection*{Outer--co-inner factorization}
For the solution of the approximate synthesis problems (i.e., AFDP, AFDIP or AMMP), the co-outer--co-inner factorization  plays an important role. We only consider the particular case, of a stable and full row rank rational matrix $G(\lambda)$, without zeros on the boundary of the stability region, which can be factored as
\be\label{oifac} G(\lambda) = [\, G_o(\lambda) \; 0 \,]  G_i(\lambda) , \ee
where $[\, G_o(\lambda) \; 0\,]$ is the co-outer factor with  $G_o(\lambda)$ invertible, stable and having only stable zeros (i.e, minimum-phase), and $G_i(\lambda)$ is inner (i.e., stable and $G_i(\lambda)G_i^\sim(\lambda) = I$; recall that $G_i^\sim(\lambda)$ is $G_i^T(-s)$ for a continuous-time system and $G_i^T(1/z)$ for a discrete-time system). If we partition $G_i(\lambda)$ compatibly with $[\, G_o(\lambda) \; 0\,]$ as $G_i(\lambda) = \left[\begin{smallmatrix} G_{i,1}(\lambda) \\ G_{i,2}(\lambda) \end{smallmatrix}\right]$, then $G(\lambda) = G_o(\lambda)G_{i,1}(\lambda)$ is the outer--co-inner factorization of $G(\lambda)$ and (\ref{oifac}) is the extended co-outer--inner factorization of $G(\lambda)$.

The role of this factorization is twofold. The post-multiplication of $G(\lambda)$ with $G_i^\sim(\lambda)$ achieves the column compression of $G(\lambda)$ to a full column rank matrix $G_o(\lambda)$, which is invertible (being also full row rank). Moreover, all resulting zeros of $G_o(\lambda)$ are stable, and therefore its inverse $G_o^{-1}(\lambda)$ is stable as well. In solving approximate synthesis problems, $G(\lambda)$ is either $R_w(\lambda)$ or $[\, R_f(\lambda)\; R_w(\lambda)\,]$ and the corresponding factor, say $Q_4(\lambda)$, is chosen $Q_4(\lambda) = G_o^{-1}(\lambda)$. If the factorizations algorithms proposed in \citep{Oara00} in the continuous-time case and in \citep{Oara05} for the discrete-time case, are employed, then the resulting outer factor has a realization of the form
\[ G_o(\lambda) = \ba{c|c} A_Q-\lambda E_Q & \widetilde B \\ \hline \\[-3mm]
C_Q & \widetilde D \ea \]
for certain $\widetilde B$ and $\widetilde D$ and explicit realizations of the updated filters $[\,Q(\lambda)\; R(\lambda)\,] \leftarrow G_o^{-1}(\lambda)[\,Q(\lambda)\; R(\lambda)\,]$
can be computed in the forms compatible with (\ref{QR}) as
\[ G_o^{-1}(\lambda)[\,Q(\lambda)\; R(\lambda)\,]  = {\arraycolsep=1mm\ba{cc|cc}  A_Q-\lambda E_Q & \widetilde B & B_Q & B_R\\ C_Q & \widetilde D & D_Q & D_R\\ \hline 0 & -I & 0 & 0\ea} . \]

 In the case when $G(\lambda)$ has zeros on the boundary of the stability domain, a factorization as in (\ref{oifac}) can still be computed, with $G_o(\lambda)$ containing, besides the stable zeros, also all zeros on the boundary of the stability domain. In this case, the inverse $G_o^{-1}(\lambda)$ is unstable, or improper or both of them. Such a case is called \emph{non-standard} and, when it is encountered, requires a special treatment. We should emphasize that non-standard cases occur only in conjunction with employing a particular solution method and are generally not related to the solvability of the fault diagnosis problems.

\section{Synthesis Procedures}
We present in this section three synthesis procedures of fault detection filters, to illustrate the application of the computational paradigms discussed in the previous section. A complete set of synthesis procedures for all formulated synthesis problems is presented in \citep{Varg17}.

As a first example, we give the complete synthesis procedure to solve the AFDP. This procedure can also be applied to solve the EFDP (if no noise inputs are present), and serves as a computational kernel for solving the AFDIP (and also of EFDIP) by computing repeatedly the solutions of appropriately formulated AFDPs.\\

\noindent\textbf{Procedure AFD}\\[2mm]
\hspace*{-2mm}\begin{tabular}{ll}\emph{Inputs: } & $\{G_u(\lambda), G_d(\lambda), G_f(\lambda), G_w(\lambda)\}$\\
\emph{Outputs: }& $Q(\lambda)$, $R_f(\lambda)$, $R_w(\lambda)$, $\eta$
\end{tabular}
\begin{enumerate}
\item Compute  a  proper minimal {left nullspace} basis $Q_1(\lambda)$ of $G(\lambda)$ in (\ref{null});
set $Q(\lambda) = Q_1(\lambda)$ and compute
\[ R(\lambda) := [\, R_f(\lambda) \; R_w(\lambda) \,]  = Q_1(\lambda){\arraycolsep=0.5mm\ba{cc}G_f(\lambda) & G_w(\lambda) \\  0 & 0\ea }. \] \\
\textbf{Exit} if $\exists\, j$ such that $R_{f_j}(\lambda) = 0$  (no solution).
\item Choose a proper $Q_2(\lambda)$ such that $Q_2(\lambda)Q(\lambda)$
 has least order and satisfies $Q_2(\lambda)R_{f_j}(\lambda) \not = 0$  $\forall \,j$ and $Q_2(\lambda)R_{w}(\lambda)$ has full row rank (admissibility); \\
update $[\, Q(\lambda) \; R(\lambda)\,] \leftarrow Q_2(\lambda)[\, Q(\lambda) \; R(\lambda)\,]$.
\item  Compute the left coprime factorization
\[ [ \,Q(\lambda) \; R(\lambda)\, ] = Q_3^{-1}(\lambda) [\, N_Q(\lambda) \; N_R(\lambda) \,] \]
such that $N_Q(\lambda)$
and $N_R(\lambda)$ have desired stable dynamics;
update  $[\, Q(\lambda) \; R(\lambda)\,] \leftarrow [\, N_Q(\lambda) \; N_R(\lambda) \,] $.
\item  Compute  the
outer--co-inner factorization
\(R_w(\lambda) = R_{wo}(\lambda)R_{wi}(\lambda), \) where
the outer factor $R_{wo}(\lambda)$ is invertible; with $Q_4(\lambda) =  R_{wo}^{-1}(\lambda)$  update
$[ \, Q(\lambda) \; R(\lambda) \, ] \leftarrow Q_4(\lambda)[ \, Q(\lambda) \; R(\lambda) \, ]$.
\item Evaluate the fault-to-noise gap $\eta$ in (\ref{nfgap}).
\end{enumerate}

This procedure illustrates the discussed synthesis paradigms, as the nullspace-based reduction performed at Step 1, the least-order synthesis in conjunction with admissibility conditions at Step 2, allocation of filter dynamics via coprime factorization techniques at Step 3 and the use of outer--co-inner factorization at Step 4 (with an optimal choice of $Q_4(\lambda)$ to maximize the fault-to-noise gap). The flexibility of the factorization-based synthesis can be illustrated by skipping some of the synthesis steps. For example, if the least order synthesis is not of interest, then $Q_2(\lambda) = I$ at Step 2, and if additionally, the resulting $Q_1(\lambda)$ at Step 1 has an already satisfactory dynamics, then $Q_3(\lambda) = I$ at Step 3. Step 4 need not be performed if no noise inputs are present (i.e., the EFDP is solved), in which case the fault-to-noise gap is set to $\eta = \infty$ at Step 5. The occurrence of the non-standard case with $Q_4(\lambda) = R_{wo}^{-1}(\lambda)$ unstable at Step 4, can be simply handled, for example, by repeating the computations at Step 3 after performing Step 4.

The second example is the synthesis procedure to solve the \emph{strict} AFDIP. A straightforward approach is to solve for each row of the given structure matrix $S$ an AFDP, for a synthesis model with redefined sets of disturbance and fault inputs. A potentially more efficient approach is to use first the nullspace-based paradigm to ensure the decoupling conditions $R_u(\lambda) = 0$, $R_d(\lambda) = 0$, and to obtain the  reduced system (\ref{systemfw}), and then solve for each row of $S$ an AFDP for a reduced system with redefined sets of disturbance and fault inputs. This second approach is the basis of the following synthesis procedure. \\

\noindent\textbf{Procedure AFDI}\\[2mm]
\hspace*{-1mm}{\tabcolsep=1mm\begin{tabular}{ll}\emph{Inputs: } & $\{G_u(\lambda), G_d(\lambda), G_f(\lambda), G_w(\lambda)\}$, $S \in \mathds{R}^{n_b\times m_f}$\\
\emph{Outputs: }& $Q^{(i)}(\lambda)$, $R_f^{(i)}(\lambda)$, $R_w^{(i)}(\lambda)$, $\eta_i$, $i = 1, \ldots, n_b$.
\end{tabular}}
\begin{enumerate}
\item Compute  a minimal {left nullspace} basis $Q(\lambda)$ of $G(\lambda)$ in (\ref{null})  and
\[ [\, R_f(\lambda) \; R_w(\lambda) \,]  = Q(\lambda){\arraycolsep=0.5mm\ba{cc}G_f(\lambda) & G_w(\lambda) \\  0 & 0\ea } , \]
such that $Q(\lambda)$, $R_f(\lambda)$ and $R_w(\lambda)$ are stable.    \\[-4mm]
\item For $\;i = 1, ..., n_b$\\[2mm]
\vspace*{-1mm}
\hspace*{3mm}\begin{minipage}[t]{65mm}
\begin{enumerate}
\item[2.1)] Form $\overline G_{d}^{(i)}(\lambda)$ from the columns $R_{f_j}(\lambda)$ 
for which $S_{ij} = 0$ and
$\overline G_{f}^{(i)}(\lambda)$ from the columns $R_{f_j}(\lambda)$  for which $S_{ij} \not= 0$.
\item[2.2)] Apply \textbf{Procedure AFD} to the system described by 
$\big\{0,\overline G_{d}^{(i)}(\lambda), \overline G_{f}^{(i)}(\lambda), R_{w}(\lambda)\big\}$
to obtain the stable filter $\overline Q^{(i)}(\lambda)$ and $\eta_i$. \\
\textbf{Exit} if no solution exists.
\item[2.3)] Compute $Q^{(i)}(\lambda) = \overline Q^{(i)}(\lambda)Q(\lambda)$,
$R_f^{(i)}(\lambda) = \overline Q^{(i)}(\lambda)R_f(\lambda)$ and $R_w^{(i)}(\lambda) = \overline Q^{(i)}(\lambda)R_w(\lambda)$.
\end{enumerate}
\end{minipage}\\
\end{enumerate}

This procedure can be easily adapted to solve a \emph{soft} AFDIP, by applying at Step 2.2 the \textbf{Procedure AFD} to the quadruple $\big\{0,0, \overline G_{f}^{(i)}(\lambda), \big[\, R_{w}(\lambda) \; \overline G_{d}^{(i)}(\lambda)\,\big]\big\}$. Moreover, \textbf{Procedure AFDI} can be also employed to solve an EFDIP in the case when there are no noise inputs.

Finally, the third example is the synthesis procedure to solve the AMMP.  The standard approach to address the solution of the AMMD is to determine a stable $Q(\lambda)$ which fulfills $R_u(\lambda) = 0$ and $R_d(\lambda) = 0$ and, simultaneously, minimizes the error norm $\|\mathcal{E}(\lambda)\|$, with
\[ \mathcal{E}(\lambda) := [\, R_f(\lambda) - M_r(\lambda) \;\;  R_w(\lambda) \,] . \]
This problem is solved by  reducing it, using factorization techniques, to a least distance problem.
Since the resulting optimal solution $Q(\lambda)$ may not result proper or stable, an updating factor $M(\lambda)$ can be used such that $M(\lambda)Q(\lambda)$ is stable. This updated filter can be interpreted as a suboptimal solution of a modified error norm minimization problem with $M_r(\lambda)$ replaced by $M(\lambda)M_r(\lambda)$.

The following synthesis procedure can be used to solve the AMMP for a strongly isolable system and an invertible and diagonal reference
model $M_r(\lambda)$ using the  error norm minimization approach sketched above.\\

\noindent\textbf{Procedure AMMS}\\[2mm]
\hspace*{-2mm}\begin{tabular}{ll}\emph{Inputs: } & $\{G_u(\lambda), G_d(\lambda), G_f(\lambda), G_w(\lambda)\}$, \\ & $M_r(\lambda)$ (invertible and diagonal)\\
\emph{Outputs: }& $Q(\lambda)$, $R_f(\lambda)$, $R_w(\lambda)$, $M(\lambda)$ (diagonal)
\end{tabular}
\begin{enumerate}
\item Compute  a proper minimal {left nullspace} basis $Q_1(\lambda)$ of $G(\lambda)$ in (\ref{null});
set $Q(\lambda) = Q_1(\lambda)$ and compute
\[ R(\lambda) := [\, R_f(\lambda) \; R_w(\lambda) \,]  = Q_1(\lambda){\arraycolsep=0.5mm\ba{cc}G_f(\lambda) & G_w(\lambda) \\  0 & 0\ea }. \] \\
\textbf{Exit} if $\rank R_{f}(\lambda) < m_f$  (no solution).
\item Choose $Q_2(\lambda)$ such that $Q_2(\lambda)R(\lambda)$ has maximal full row rank  and $Q_2(\lambda)Q(\lambda)$ has least McMillan degree; update $[ Q(\lambda) \; R(\lambda)] \leftarrow Q_2(\lambda)[\, Q(\lambda) \; R(\lambda)\,]$.
\item Compute the extended co-outer--inner factorization
\[ R(\lambda) = [\, R_{o}(\lambda) \;  0 \, ] \ba{c} R_{i,1}(\lambda) \\ R_{i,2}(\lambda) \ea .\]
With $Q_3(\lambda) =  R_{o}^{-1}(\lambda)$,
 update $[ Q(\lambda) \; R(\lambda)] \leftarrow Q_3(\lambda)[\, Q(\lambda) \; R(\lambda)\,]$ and compute
 \[ [\,  F_1(\lambda) \;  F_2(\lambda) \,] := [M_r(\lambda) \; 0][\, R_{i,1}^\sim(\lambda) \; R_{i,2}^\sim(\lambda) \,] .\]
\item Compute the stable solution $Q_4(\lambda)$ of the least-distance problem
\[\min_{Q_4(\lambda) \in \mathcal{H}_\infty} \left\|{\arraycolsep=1mm\ba{cc} F_1 (\lambda)- Q_4(\lambda) &  F_2(\lambda)  \ea } \right\|; \] 
    update $[\, Q(\lambda) \; R(\lambda)\,] \leftarrow Q_4(\lambda)[\, Q(\lambda) \; R(\lambda)\,]$.
\item  Determine diagonal, invertible and stable  $Q_5(\lambda) := M(\lambda)$ such that $M(\lambda)[\, Q(\lambda) \; R(\lambda)\,]$ is stable; update $[\, Q(\lambda) \; R(\lambda)\,] \leftarrow Q_5(\lambda)[\, Q(\lambda) \; R(\lambda)\,]$.
\end{enumerate}

The solution of the least-distance problem at Step 4 depends on the employed norm  (i.e., either the $\mathcal{H}_2$ norm or $\mathcal{H}_\infty$ norm) (see, for example, \citep{Varg17} for suitable computational procedures).

The Procedure AMMS can be also employed to solve an EMMP for a strongly isolable system, with obvious simplifications (e.g., using $Q_3(\lambda) = R_f^{-1}(\lambda)$ at Step 3 and $Q_4(\lambda) = M_r(\lambda)$ at Step 4).

\section{Software Tools}
The development of dedicated software tools for the synthesis of fault detection filters must fulfill some general requirements for robust software implementation, but also  some requirements specific to the field of fault detection. In what follows, we shortly discuss some of these requirements.

A basic requirement is the use of general, numerically reliable and computationally efficient numerical approaches as basis for the implementation of all computational functions, to guarantee the solvability of problems under the most general existence conditions of the solutions. Synthesis procedures suitable for robust software implementations are described in the book \citep{Varg17}.

For beginners, it is important to be able to easily obtain preliminary synthesis results. Therefore, for ease of use, simple and uniform user interfaces are desirable for all synthesis functions. This can be achieved by relying on meaningful default settings for all problem parameters and synthesis options.  Also, the synthesis functions to solve approximate synthesis problems should be applicable to solve the exact synthesis problems as well. On the other side, the solution of an exact problem for a system with noise inputs, should provide a first approximation to the solution of the approximate synthesis problem.

For experienced users, an important requirement is to provide an exhaustive set of options to ensure the complete freedom in choosing problem-specific parameters and synthesis options. Among the frequently used synthesis options we mention: the number of residual signal outputs; stability degree for the poles of the resulting filters or the location of their poles; type of the employed nullspace basis (e.g., minimal proper, full-order observer based); performing least-order synthesis, etc.

The \emph{Fault Detection and Isolation Tools} \linebreak[4](FDITOOLS) is a publicly available collection of {MATLAB} \emph{m}-files for the analysis and solution of fault diagnosis problems and has been implemented along the previously formulated requirements. FDITOOLS
supports the basic  synthesis approaches of linear residual generation filters for both continuous-time and discrete-time LTI systems. The underlying synthesis techniques rely on reliable numerical algorithms developed by the author and are described in the Chapters 5--7 of the  book \citep{Varg17}.
The functions of the FDITOOLS collection rely on the \emph{Control System Toolbox} \citep[and later releases]{MLCO15} and the \emph{Descriptor System Tools} ({DSTOOLS}) \citep{Varg18}, which provides the basic tools for the manipulation of rational TFMs via their descriptor system representations.
The version V1.0 of the {FDITOOLS} collection covers all synthesis procedures described in  \citep{Varg17} and, additionally, includes several useful analysis functions,  as well as functions for an easy setup of synthesis models.
A precursor of FDITOOLS was the {Fault Detection} Toolbox for MATLAB, a  proprietary software of the German Aerospace Center (DLR), developed between 2004 and 2014 (for the status of this toolbox around 2006 see \citep{Varg06c}).

A notable recent development is the free software package \textbf{\texttt{FaultDetectionTools}}\footnote{\textbf{\url{https://github.com/andreasvarga/FaultDetectionTools.jl}}},  which implements the complete functionality of FDITOOLS in the Julia language. Julia is a powerful and flexible dynamic language, suitable for scientific and numerical computing, with performance comparable to traditional statically-typed languages such as Fortran or C. As a programming  language, Julia features optional typing, multiple dispatch, and good performance, achieved using type inference and just-in-time compilation \citep{Bezanson2017}. The  underlying Julia packages  \textbf{\texttt{DescriptorSystems}}\footnote{\textbf{\url{https://github.com/andreasvarga/DescriptorSystems.jl}}},  for handling descriptor system representations, \textbf{\texttt{MatrixEquations}}\footnote{\textbf{\url{https://github.com/andreasvarga/MatrixEquations.jl}}}, for solving  various control related matrix equations (Lyapunov, Sylvester, Riccati), and \textbf{\texttt{MatrixPencils}}\footnote{\textbf{\url{https://github.com/andreasvarga/MatrixPencils.jl}}}, for manipulation of matrix pencils, 
provide the required basic computational functionality for the implementation of \textbf{\texttt{FaultDetectionTools}}.   

\section{Recommended Reading}
The book \citep{Varg17} provides extensive information  on the mathematical background of solving synthesis problems of fault detection filters, gives detailed descriptions of the underlying synthesis procedures and contains an extensive list of references to the relevant literature. The numerical aspects of the synthesis procedures are also amply described (a novelty in the fault detection related literature).  Additionally, this book addresses the related model detection problematic, a multiple-model based approach to solve fault diagnosis problems (e.g., for parametric or multiplicative faults). For a concise presentation of the descriptor system techniques employed in the synthesis procedures, see the companion article \citep{Varg19}. A shorter version of this article appeared in the  Encyclopedia of Systems and Control \citep{Varg19e}.
FDITOOLS is distributed as a free software via the Bitbucket repository \textbf{\url{https://bitbucket.org/DSVarga/fditools}}. A comprehensive documentation of version V1.0 is available in arXiv \citep{Varg18a}. \\[-7mm]

%


\begin{thebibliography}{15}
\providecommand{\natexlab}[1]{#1}
\providecommand{\url}[1]{{#1}}
\providecommand{\urlprefix}{URL }
\expandafter\ifx\csname urlstyle\endcsname\relax
  \providecommand{\doi}[1]{DOI~\discretionary{}{}{}#1}\else
  \providecommand{\doi}{DOI~\discretionary{}{}{}\begingroup
  \urlstyle{rm}\Url}\fi
\providecommand{\eprint}[2][]{\url{#2}}

\bibitem[Bezanson et~al.(2017)]{Bezanson2017}
Bezanson J, Edelman A, Karpinski S,  Shah VB
(2017): Julia: A fresh approach to numerical computing.
SIAM Review, vol 59, 1:65--98
see also: \url{https://julialang.org/}

\bibitem[{Blanke et~al(2016)Blanke, Kinnaert, Lunze, and Staroswiecki}]{Blan16}
Blanke M, Kinnaert M, Lunze J, Staroswiecki M (2016) Diagnosis and
  Fault-Tolerant Control, 3rd ed. Springer-Verlag, Heidelberg

\bibitem[{Chen and Patton(1999)}]{Chen99}
Chen J, Patton RJ (1999) Robust Model-Based Fault Diagnosis for Dynamic
  Systems. Kluwer Academic Publishers, London

\bibitem[{Ding(2013)}]{Ding13}
Ding SX (2013) Model-based Fault Diagnosis Techniques, 2nd Edition. Springer
  Verlag, Berlin

\bibitem[{Gertler(1998)}]{Gert98}
Gertler J (1998) Fault Detection and Diagnosis in Engineering Systems. Marcel
  Dekker, New York

\bibitem[{Isermann(2006)}]{Iser06}
Isermann R (2006) Fault-Diagnosis Systems, An Introduction from Fault Detection
  to Fault Tolerance. Springer, Berlin

\bibitem[{MathWorks(2015)}]{MLCO15}
MathWorks (2015) Control System Toolbox (R2015b), User's Guide. The MathWorks
  Inc., Natick, MA

\bibitem[{Oar\u{a}(2005)}]{Oara05}
Oar\u{a} C (2005) Constructive solutions to spectral and inner-outer
  factorizations with respect to the disk. Automatica 41:1855--1866,
  \doi{10.1016/j.automatica.2005.04.009}

\bibitem[{Oar\u{a} and Varga(2000)}]{Oara00}
Oar\u{a} C, Varga A (2000) Computation of general inner-outer and spectral
  factorizations. IEEE Trans Automat Control 45:2307--2325

\bibitem[{Varga(2006)}]{Varg06c}
Varga A (2006) A {\textsc{fault detection}} toolbox for \textsc{Matlab}. In:
  Proceedings of the IEEE Conference on Computer Aided Control System Design,
  Munich, Germany, pp 3013--3018

\bibitem[{Varga(2017{\natexlab{a}})}]{Varg17d}
Varga A (2017{\natexlab{a}}) On recursive computation of coprime factorizations
  of rational matrices. \url{https://arxiv.org/abs/1703.07307}

\bibitem[{Varga(2017{\natexlab{b}})}]{Varg17}
Varga A (2017{\natexlab{b}}) Solving Fault Diagnosis Problems -- Linear
  Synthesis Techniques, Studies in Systems, Decision and Control, vol~84.
  Springer International Publishing, \doi{10.1007/978-3-319-51559-5}

\bibitem[{Varga(2018{\natexlab{a}})}]{Varg18}
Varga A (2018{\natexlab{a}}) {Descriptor System Tools (DSTOOLS) V0.71, User's
  Guide}. \url{https://arxiv.org/abs/1707.07140}

\bibitem[{Varga(2018{\natexlab{b}})}]{Varg18a}
Varga A (2018{\natexlab{b}}) {Fault Detection and Isolation Tools (FDITOOLS)
  V1.0, User's Guide}. \url{https://arxiv.org/abs/1703.08480}

\bibitem[{Varga(2019{\natexlab{a}})}]{Varg19}
Varga A (2019{\natexlab{a}}) {Descriptor system techniques and software tools}.
  \url{https://arxiv.org/abs/1902.00009}

\bibitem[{Varga(2019{\natexlab{b}})}]{Varg19e}
Varga A (2019{\natexlab{b}}) Fault detection and diagnosis: Computational
  issues and tools. In: Baillieul J, Samad T (eds) Encyclopedia of Systems and
  Control, Springer London,
  \doi{10.1007/978-1-4471-5102-9_100055-1},
  \urlprefix\url{https://doi.org/10.1007/978-1-4471-5102-9_100055-1}

\end{thebibliography}

\end{document}